\documentclass[10pt, a4paper, onecolumn]{scrartcl}

\usepackage{graphicx}
\usepackage[format=plain]{caption}
\usepackage[left=1.6cm,right=1.6cm,top=1cm,bottom=1cm, headheight=15pt, headsep=1cm, includeheadfoot]{geometry}
\usepackage[onehalfspacing]{setspace}

\begin{document}

\begin{center}
	\huge{\noindent \textbf{Ultrafast, all-optical coherent control of single silicon vacancy colour centres in diamond}}\\
	\quad\\
	\normalsize
	Jonas Nils Becker$^1$, Johannes G\"orlitz$^1$, Carsten Arend$^1$, Matthew Markham$^2$ and Christoph Becher$^{1,*}$\\
	\quad\\
	\small{
	$^1$Fachrichtung 7.2 (Experimentalphysik), Universit\"at des Saarlandes,  Campus E2.6, 66123 Saarbr\"ucken, Germany\\
	$^2$Element Six Limited, Global Innovation Centre, Fermi Avenue, Harwell Oxford, Didcot, Oxfordshire OX11 0QR, United Kingdom}
	\normalsize
	\quad\\\quad\\\quad\\
\end{center}

\begin{figure}[!b] 
	\centering
	\includegraphics[width=0.5\columnwidth]{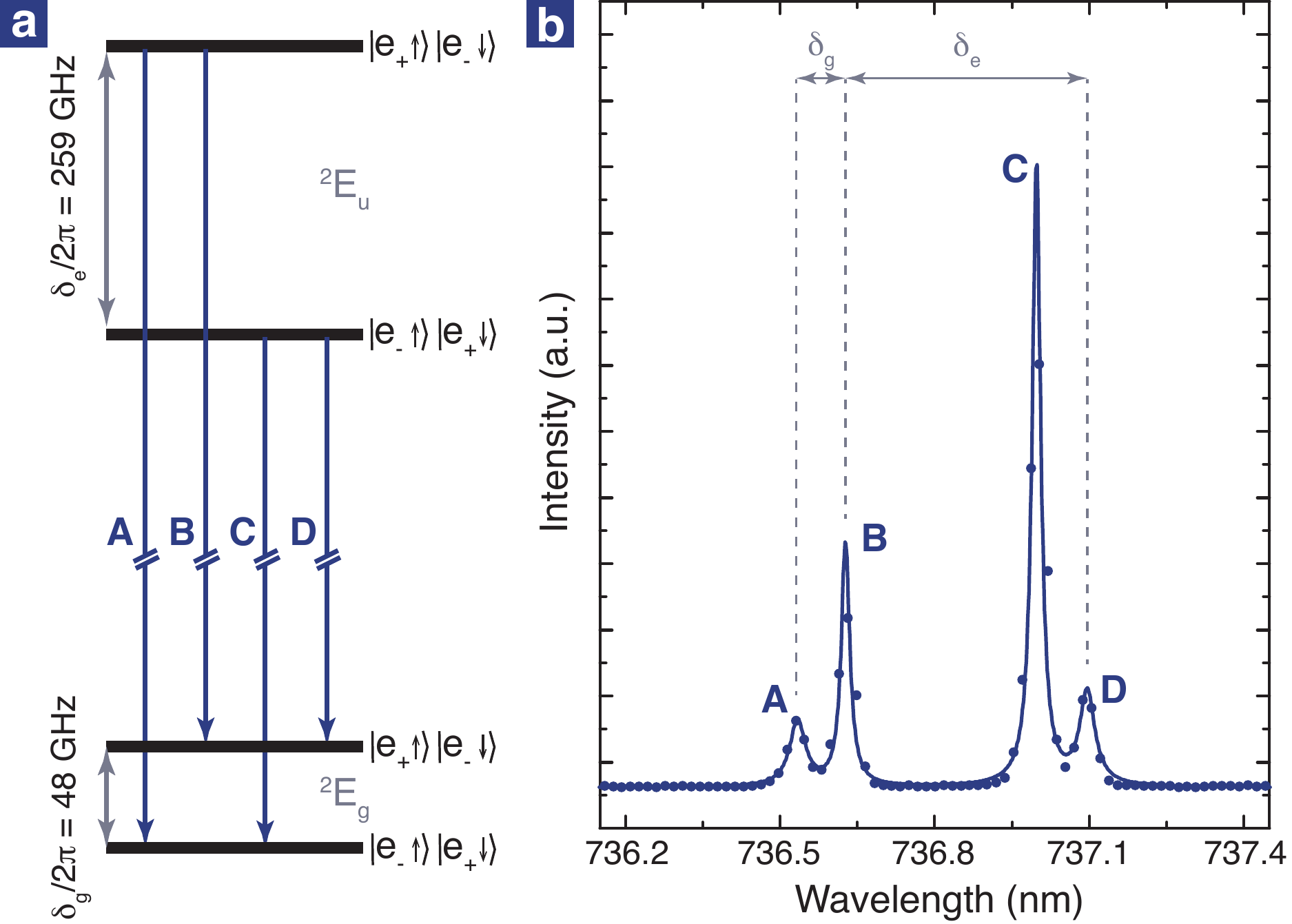}
	\caption{{\footnotesize \textbf{Spectrum and Electronic Structure of the SiV$^-$.} (a) Electronic level scheme of the SiV$^-$ at zero magnetic field. (b) Photoluminescence Spectrum of the SiV$^-$ used throughout this study. The values of the spectral splitting between transitions A and B as well as A and C indicate an unstrained centre.}}
	\label{fig:Fig1} 
\end{figure}

\begin{figure}[!b] 
	\centering
	\includegraphics[width=0.5\columnwidth]{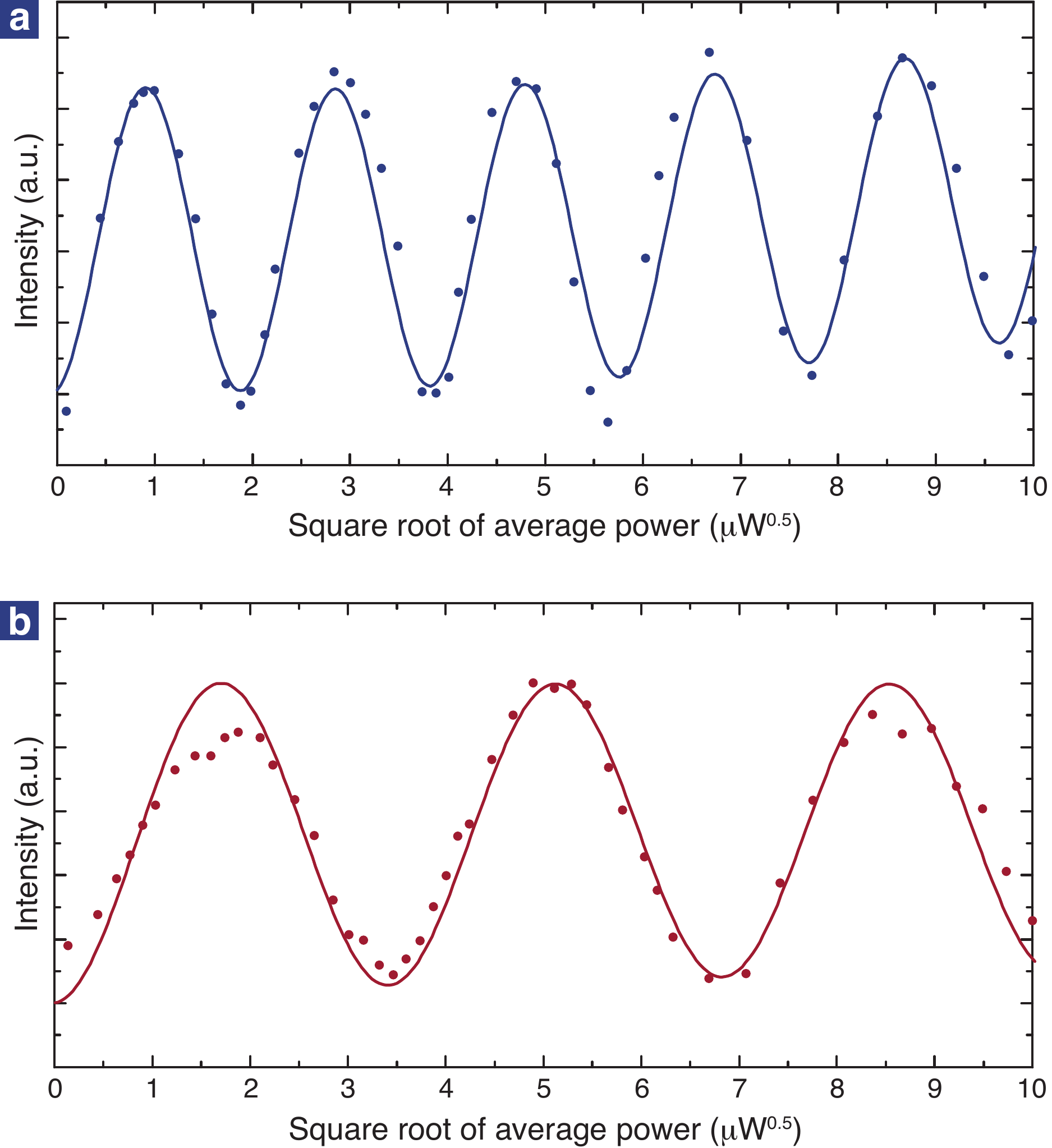}
	\caption{{\footnotesize \textbf{Optical one-photon Rabi oscillations.} Measured photon counts as a function of average laser power (after subtraction of background fluorescence) for the laser resonant with (a) transition C and (b) transition B. In both cases pulses with 12\,ps length and a double-sided exponential temporal shape (due to filtering with a Fabry-Perot etalon) have been applied. The data in both graphs is modelled by a four-level density matrix model (solid lines) including spontaneous decay between the excited and ground states as well as the phonon-induced decay and thermalization processes within the ground and excited state manifolds. All rates have been measured experimentally and no additional free parameters have been employed to model the Rabi oscillations (see supplemental material)\cite{supplement}.
		}}
		\label{fig:Fig2}
\end{figure}

\begin{figure}[!t] 
	\centering
	\includegraphics[width=0.5\columnwidth]{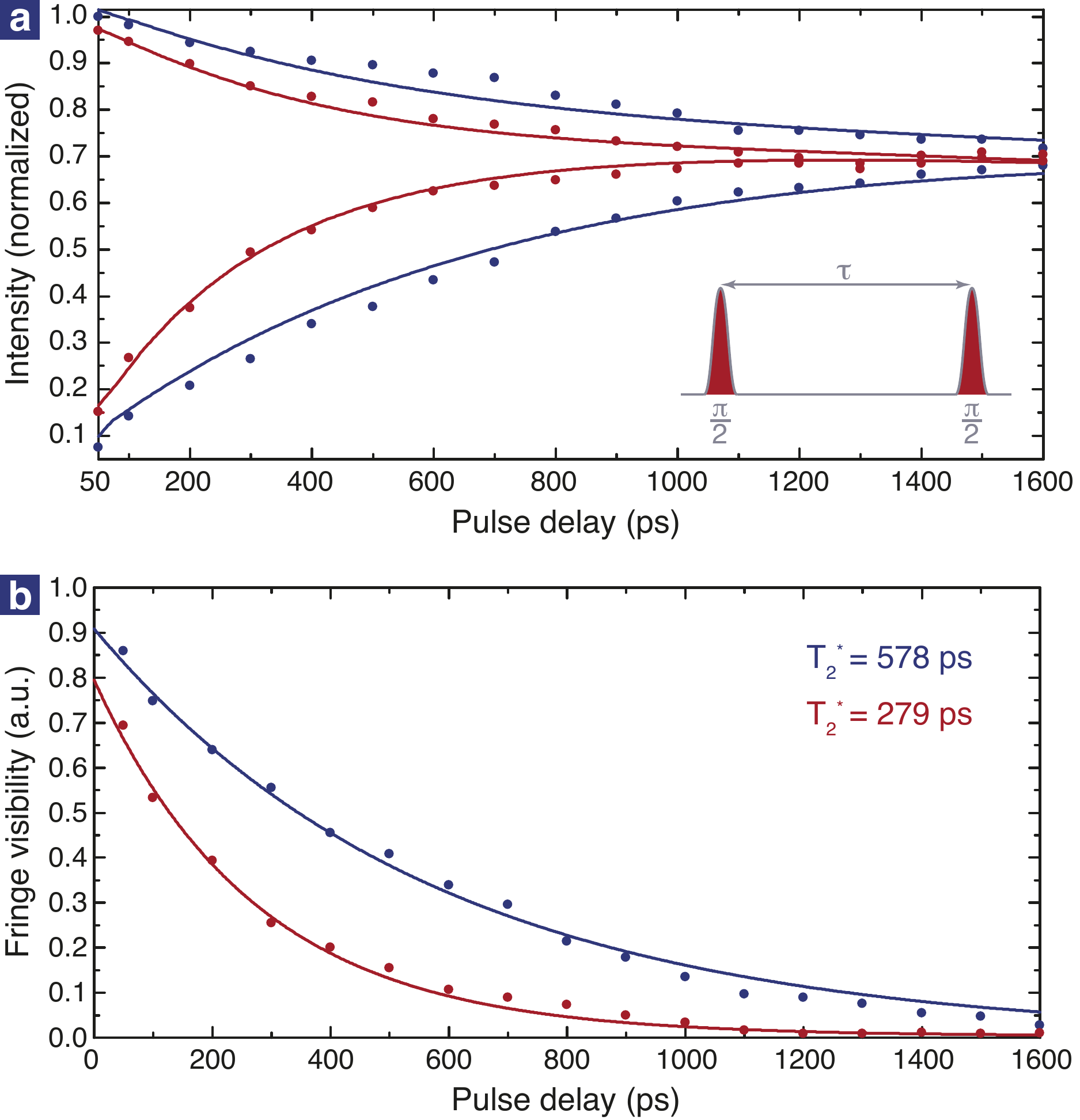}
	\caption{{\footnotesize \textbf{Optical one-photon Ramsey interference.} Interference fringes can be observed by applying a sequence of two picosecond $\frac{\pi}{2}$ pulses with variable delay. (a) Upper and lower envelopes of the observed interference pattern for transition C (blue) and transition B (red). (b) Fringe visibilities and resulting excited state coherence times for transition C (blue) and B (red).}}
	\label{fig:Fig3}
\end{figure}

\begin{figure}[!b] 
	\centering
	\includegraphics[width=0.5\columnwidth]{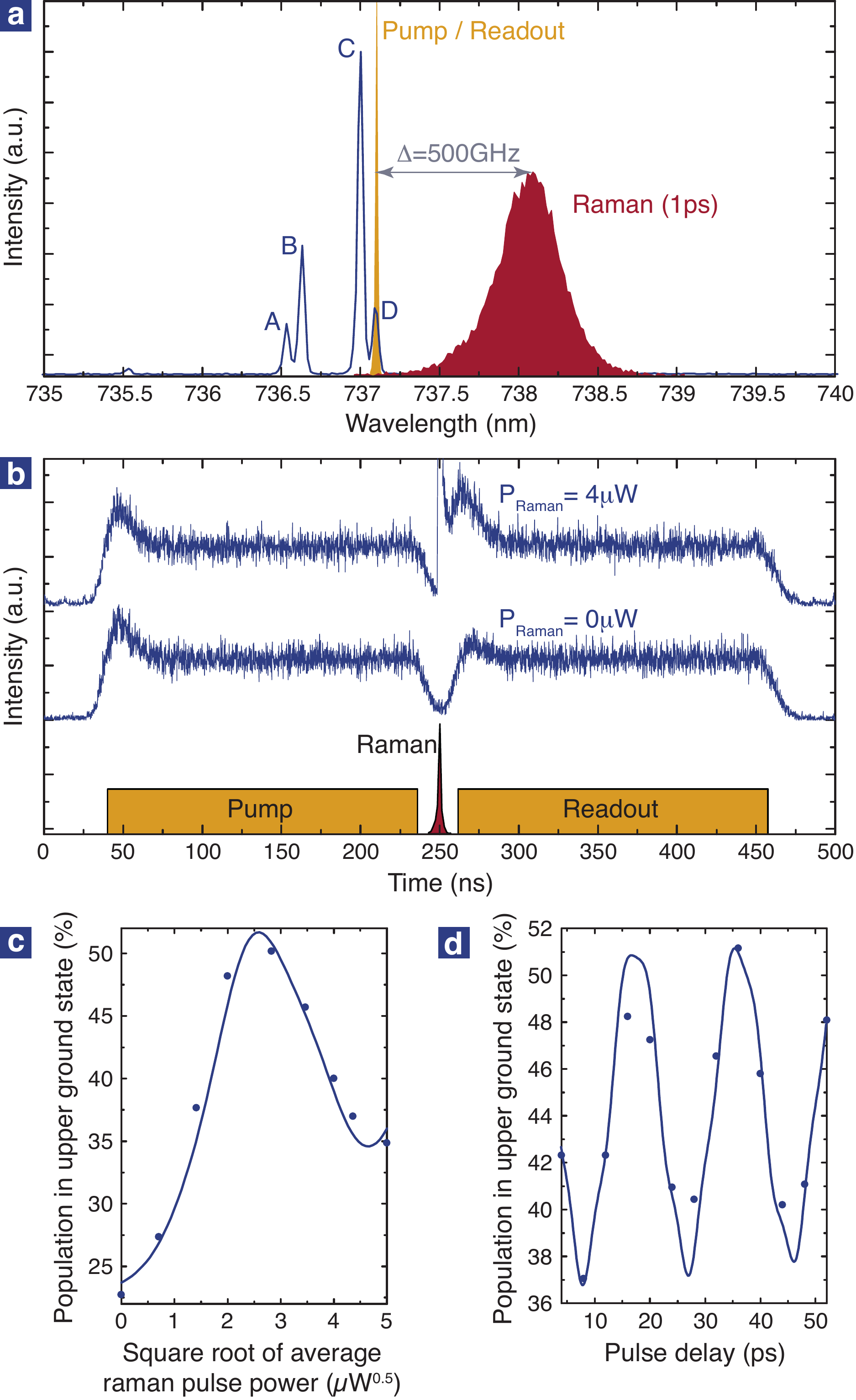}
	\caption{{\footnotesize \textbf{Raman-based population transfer between ground states.} (a) Spectrum displaying spectral position of the SiV$^-$ as well as width and detuning of the CW Pump/Readout and pulsed Raman rotation laser involved. (b) Schematic pulse sequence and measured fluorescence response for Raman beam switched off and P$_{\rm Raman}$=4\,$\mu$W (red). (c) Transferred population as a function of the average power in the Raman beam. (d) Ramsey interference generated by two subsequent Raman pulses verifying coherent transfer. Solid lines show simulations using the four-level density matrix model with the relative driving strength of both Raman transitions as the only free parameter. The model indicates that the ratio of the driving strengths of transitions C and D in the Raman beam is about 1\,:\,0.7 after optimizing the polarization.}}
	\label{fig:Fig4}
\end{figure}

\textbf{Complete control of the state of a quantum bit (qubit) is a fundamental requirement for any quantum information processing (QIP) system. For this purpose, all-optical control techniques offer the advantage of a well-localized and potentially ultrafast manipulation of individual qubits in multi-qubit systems. Fast manipulation, however, requires qubits with large electronic level splittings to enable the use of broadband laser pulses. For qubits in the solid state, which are advantageous due to their good scalability, only semiconductor quantum dots fulfil these conditions so far \cite{press2008}. However, to achieve a $\Lambda$-type electronic level configuration with a large enough level splitting, charged dots in strong magnetic fields have to be used \cite{lagoudakis2013} and typical electron spin coherence times are on the order of one nanosecond \cite{gao2012,sun2016} without employing nuclear field locking techniques or spin echo sequences. Recently, the negatively charged silicon vacancy centre (SiV$^-$) in diamond has emerged as a novel promising system for QIP due to its superior spectral properties \cite{neu2011} and advantageous electronic structure \cite{hepp2014,rogers2014}: The  SiV$^-$ offers an optically accessible $\Lambda$-type level structure with a large orbital level splitting even without the need of an external magnetic field and previous studies determined the ground state coherence time of the centre to be on the order of 35-45\,ns \cite{pingault2014,rogers20142}. While these studies already indicate the feasibility of ultrafast all-optical coherent control of the SiV$^-$, no experimental realization has been presented so far. Here, we report on optical Rabi oscillations and Ramsey interference between the ground and the excited state of the SiV$^-$  using 12\,ps long, resonant rotation pulses. Moreover, utilizing a $\Lambda$-scheme, we demonstrate Raman-based coherent Rabi rotations and Ramsey interference within the ground state manifold using a single, 1\,ps long off-resonant pulse. These measurements prove the accessibility of the complete set of single-qubit operations relying solely on optical fields and pave the way for high-speed QIP applications using SiV$^-$ centres.}\\
Confined spin impurities, known as colour centres, in a spin-free diamond host lattice have proven to be promising systems for QIP as well as sensing applications, with the negatively charged nitrogen vacancy centre (NV$^-$) being the most prominent example \cite{awschalom2013,balasubramanian2008}. Over the past couple of years the SiV$^-$ emerged as a steadily growing competitor to the NV$^-$ due to its superior spectral properties such as intense, narrow zero phonon line (ZPL) emission at ambient and cryogenic temperatures as well as small Huang-Rhys factors down to S=0.08 \cite{neu2011,gorokhovsky1995}. The SiV$^-$ features a unique molecular structure with an interstitial silicon atom in between two unoccupied carbon sites. The resulting inversion symmetry of this split-vacancy configuration is responsible for the high spectral stability of the SiV$^-$ by suppressing first-order Stark shifts and thus allows for the generation of indistinguishable photons from separate SiV$^-$ centres \cite{sipahigil2014}. In previous studies \cite{hepp2014,rogers2014}, the electronic structure of the defect has been investigated in great detail revealing an S=$\frac{1}{2}$ system. Due to a strong spin orbit coupling \cite{hepp2014}, the centre features twofold orbitally split and spin degenerate ground and excited states at zero magnetic field (Fig.\,\ref{fig:Fig1}a) leading to the characteristic four-line fine structure in the ZPL emission spectrum at 737\,nm (Fig.\,\ref{fig:Fig1}b). These orbital states can be used to construct an SiV$^-$ based qubit even without the need of an external magnetic field and the large splitting of $\delta_g/2\pi$=48\,GHz in the ground and $\delta_e/2\pi$=259\,GHz in the excited state allows for ultrafast optical coherent control. The ground state coherence time $T_2^*$ of the SiV$^-$ is limited by the orbital relaxation time $T_1^{\rm Orbit}$=35\,ns \cite{rogers20142}, which results from a single-phonon vibronic process causing transitions between the orbital components of the ground state manifold \cite{jahnke2015}. Hence, by eliminating these processes in future devices coherence times can be further extended. This can be achieved by either cooling the samples below 1\,K or by manipulating the phononic environment of the centre using phononic nanostructures \cite{jahnke2015}. The rapid development of nanofabrication \cite{riedrich2014,evans2015} and growth techniques \cite{jantzen2016} potentially allows for the fabrication of diamond based phononic band gap materials \cite{kipfstuhl2014} or small nanodiamonds to suppress ground state thermalization. As a result, coherence times in the millisecond range seem feasible as the pure spin relaxation time has been determined to be $T_1^{\rm Spin}$=2.4\,ms \cite{rogers20142}.\\
A universal single-qubit gate requires control of the angle (U(1) control) as well as the axis of rotation (SU(2) control) of a qubit state. First, we demonstrate resonant angular control by coherently driving Rabi oscillations between the ground and the excited state of the centre using 12\,ps laser pulses. Figure \,\ref{fig:Fig2}a and b show the fluorescence signal from the excited state for variable pulse areas of the picosecond laser resonant with transitions C and B (see Fig.\,\ref{fig:Fig1}). High contrast Rabi oscillations are evident in the photon count rate traces of both transitions with visibilities exceeding 90\,\% and are well-fitted by a four level density matrix model (solid lines) \cite{supplement}. Coherent rotations up to $\Theta\approx10\pi$ for transition C and up to $\Theta\approx6\pi$ for transition B are observed without any significant damping, with the rotation angles being limited by the available laser power. Residual background fluorescence of the diamond causes slight upwards slopes in both curves.\\
The oscillations shown in Fig.\,\ref{fig:Fig2} demonstrate rotation about the x axis of the Bloch sphere. However, full SU(2) control requires rotation around a second axis. This can be achieved by exploiting the free precession of the state about the vertical z axis of the Bloch sphere by employing a Ramsey-type pulse sequence consisting of two $\frac{\pi}{2}$ pulses separated by a variable delay time. As the experimental resonance frequency is about 406.8\,THz, the oscillations in the signal are extremely fast and no single Ramsey fringes have been resolved. However, by fine-tuning the temporal spacing of the two pulses for a number of  fixed delays it is possible to determine the maximum and minimum count rate at each fixed delay point and therefore to measure the upper and lower envelopes of the Ramsey curve which are shown in Fig.\,\ref{fig:Fig3}a for transitions B and C. Simulations (solid lines) using the four-level model show that two effects are responsible for the decrease of the Ramsey visibilities in Fig.\,\ref{fig:Fig3}b \cite{supplement}. First, spontaneous decay and thermalization effects lead to a decoherence of the system. Secondly, a pure dephasing of the excited states of about $\gamma/2\pi$=160\,MHz, most likely caused by residual phonon broadening, has to be added to fit the observed decrease in visibility. Calculations of the excited state coherence times from these visibilities yield $T_2^*$=578\,ps for the lower and $T_2^*$=279\,ps for the upper excited state, with the latter one decohering faster due to a rapid decay into the lower state. The resulting linewidths $\Delta\nu=(2\pi T_2^*)^{-1}$ of 275\,MHz for C and 570\,MHz for B are in very good agreement with the linewidths measured using photoluminescence excitation spectroscopy (270\,MHz and 574\,MHz) indicating that coherently driving the SiV$^-$, even with intense pulsed fields, does not alter its coherence properties.\\
The resonant control techniques presented so far require excitation of the SiV$^-$ and hence coherence times of such a qubit are ultimately limited by the excited state dynamics. To harness the longer ground state coherence time of the SiV$^-$, coherent rotations solely within the ground state manifold are desirable. This can be achieved by driving a $\Lambda$-type level scheme between the two ground and a common excited state. A Raman process then leads to a coherent transfer of population between the ground states. To minimize an unwanted population of the excited state, both arms of the $\Lambda$-scheme are far-detuned from one-photon resonance while still fulfilling a two-photon resonance condition. We here simultaneously couple both transitions of the $\Lambda$-system employing a single broadband laser pulse with a length of only 1\,ps, red-detuned by $\Delta$=500\,GHz from the lower excited state \cite{press2008,fu2008}. Furthermore, initialization of the population into the lower ground state and readout is achieved by optical pumping of transition D using resonant 200\,ns long laser pulses \cite{supplement}. After pumping the population in the upper ground state reaches a minimum value of 22\%. The spectral arrangement of the two lasers relative to the SiV$^-$ spectrum is depicted in Fig.\,\ref{fig:Fig4}a and the pulse sequence as well as two count rate traces at two different Raman pulse powers are shown in Fig.\,\ref{fig:Fig4}b. As the two branches of the $\Lambda$-system have different transition dipole moments and polarizations, driving strengths of both Raman transitions have to be equalized to maximize transfer efficiency. This is carried out by rotating the polarization of the Raman beam at a fixed pulse power. At the optimized polarization, the population of the upper ground state is measured for varying Raman pulse areas. The resulting curve is depicted in Fig.\,\ref{fig:Fig4}c, showing a two-photon Rabi oscillation with rotation angles of up to $\Theta\approx2\pi$, limited by the rotation laser power. As before, the data is fitted well by our model (solid curve) with the same set of experimentally measured parameters that have been used for the one-photon experiments and with the relative driving strength of the two Raman transitions being the only free parameter. The model also includes transfer via the second $\Lambda$-scheme between the ground states and the upper excited state, from which the laser is detuned by $\Delta+\delta_e=759$\,GHz. Simulations indicate that the decrease in visibility for large angles is not due to ground state decoherence but due to slight resonant excitation of the excited states for large pulse areas. In future experiments this can be minimized by modifying the laser system to provide longer pulses with a duration of a few ps. To demonstrate full SU(2) control of the ground state, also a Ramsey-type experiment has been performed and the resulting data is shown in Fig.\,\ref{fig:Fig4}d. Oscillations at the frequency of the ground state splitting of $\delta_g/2\pi$=48\,GHz are clearly visible and are reproduced well by the model (solid line), without any additional free parameters. The slight distortion of the curve is again due to resonant background fluorescence and the population does not reach the optically pumped minimum as a small amount of incoherent transfer occurs \cite{supplement,fu2008}. To resolve both issues in future experiments more complex transfer schemes with two separate rotation fields can be employed, that avoid resonant excitation and allow for more accurate optimization of the relative transition strengths. Moreover, a time delay between both fields can be used to optimize the adiabaticity of the process \cite{bergmann2015}.\\
In conclusion we demonstrated resonant and off-resonant Raman-based coherent control of the rotation angle and axis of the orbital degree of freedom of an SiV$^-$ employing ultrafast rotation pulses as short as 1\,ps. All the experiments in this work have been carried out on the orbital states of the SiV$^-$ without the need of a strong magnetic field. However, we'd like to emphasize that the techniques presented here can directly be applied to the spin states of the SiV$^-$. Although the coherence time is reduced for intermediate field strengths, we demonstrated that it recovers in the high field regime \cite{pingault2014}. This regime is required in any case to achieve large Zeeman splittings to enable the use of ultrashort pulses. The results presented here provide the basis for a number of exciting QIP applications with the SiV$^-$ such as cavity-assisted Raman transfer schemes \cite{cirac1997,hennrich2000}, coherent spin-photon interfaces \cite{togan2010}, optical quantum memories \cite{lvovsky2009, bussieres2013} and quantum gates based on geometrical phase acquisition \cite{yale2016}.

\section*{Methods}
{\small The sample consists of a type IIa high-pressure-high-temperature (HPHT) diamond with a cleaved (111) main surface. The sample is implanted with a dose of $10^9$ ions per cm$^2$ of $^{28}$Si$^+$ ions at an energy of 900\,keV, resulting in an approximately 50\,nm thick layer of SiV$^-$ centres 500\,nm below the diamond surface. After implantation the sample is annealed for 3\,h at 1000$^\circ$C in vacuum and for 1\,h at 465$^\circ$C in air and cleaned in peroxomonosulfuric acid to remove graphite residues. To further enhance the collection efficiency from the sample solid immersion lenses with a diameter of 1$\mu$m are milled into the diamond surface using focused ion beam (FIB) milling. To cure the host lattice damage induced by the FIB milling process the sample is annealed a second time using the above-mentioned protocol. The resulting colour centre density is approximately 0.2-0.4 SiV$^-$ per $\mu$m$^2$ allowing the investigation of single centres in a homebuilt confocal microscope equipped with a liquid helium flow cryostat and an NA=0.9 objective. All investigations have been carried out at a temperature of 5\,K. For the ultrafast control experiments a tunable mode-locked Ti:sapphire laser provides hyperbolic secant pulses with a length of 1\,ps at a repetition rate of 80\,MHz. The ground state Raman transfer is carried out without further filtering at a wavelength of 738\,nm for the Raman pulse. For the resonant picosecond control experiments the bandwidth of these pulses has to be reduced to avoid excitation of multiple transitions. Therefore the pulses are filtered using a monolithic Fabry-Perot etalon (F=50, FSR=1000\,GHz, $\Delta\nu$=20\,GHz). This results in Lorentzian-shaped pulses with a length of 12\,ps suitable to selectively excite the fine structure transitions of the SiV$^-$. Pulse intensities are stabilized using a feedback loop formed by a combination of a voice coil steering mirror in front of a single mode optical fibre and a photodiode after the fibre connected to a PID controller.
For the Raman transfer experiments an initialization of the population into the lowest ground state as well as readout of the population in the upper ground state after the transfer is necessary. This is achieved by excitation of transition D using a resonant continuous wave extended cavity diode laser (ECDL) stabilized to a wavemeter. Initialization and readout pulses are generated from this laser using an acousto-optic intensity modulator in single-pass configuration. To generate a sufficiently long time interval for initialization and readout the repetition rate of the mode-locked laser has been reduced to 1\,MHz using an electro-optic pulse picker cell consisting of a Pockels cell in between two Glan-Taylor prisms providing a pulse suppression greater than 1000 when the cell is switched off.
State detection is carried out by detecting the light emitted from the SiV$^-$ phonon sideband using a single photon detector attached to a fast counting electronics. The trigger signal provided by the pulsed laser or the pulse picker cell, respectively, is used to synchronize the picosecond laser pulses with the pump and readout pulses from the ECDL as well as with the counting electronics.
The data in the one-photon experiments is acquired by directly measuring the photon count rate averaged over 500\,ms per data point. In the two-photon experiments, time-resolved fluorescence count rate traces for the entire sequence duration of 1\,$\mu$s have been recorded with a total integration time of 200\,s per trace. The population of the upper ground state then is extracted from the rising edge of the fluorescence signal of the readout pulse.}

\section*{Acknowledgements}
This research has been partially funded by the European Community's Seventh Framework Programme (FP7/2007-2013) under Grant Agreement No.\,611143 (DIADEMS). Ion implantation was performed at and supported by \mbox{RUBION}, the central unit of the Ruhr-Universit\"at Bochum. We thank D. Rogalla for the implantation, C. Pauly for the fabrication of solid immersion lenses as well as M. Bock for technical assistance with the picosecond laser system. Moreover, we thank M. Atat\"ure, B. Pingault, G. Morigi, J. Eschner and E. Neu for helpful discussions throughout all stages of this work.\\

\section*{Author contributions}
J.N.B. designed, built and conducted the experiment with help from C.A. Theoretical simulations and analysis of experimental results have been carried out by J.N.B and J.G. The sample has been provided by M.M. The manuscript has been written by J.B., J.G. and C.B. C.B. conceived the experiment. All authors discussed the results and commented on the manuscript.\\

\end{document}